\documentclass[twocolumn]{aastex631}

\usepackage{graphicx}
\usepackage{amsmath}
\usepackage[utf8x]{inputenc}
\usepackage[encapsulated]{CJK}
\usepackage{ucs}
\usepackage{chngcntr}
\graphicspath{{./}{figs/}}
\usepackage[normalem]{ulem}

\newcommand{\thirteen}{UNCOVER-z13}
\newcommand{\twelve}{UNCOVER-z12}
\newcommand{\zspecth}{$z_{\rm spec} = 13.079^{+0.013}_{-0.001}$}
\newcommand{\zspectw}{$z_{\rm spec} = 12.393^{+0.004}_{-0.001}$}

\newcommand{\msae}{\texttt{msaexp}}

\newcommand{\eazy}{\texttt{EAzY}}
\newcommand{\prospector}{\texttt{Prospector}}
\newcommand{\bagpipes}{\texttt{Bagpipes}}
\newcommand{\beagle}{\texttt{BEAGLE}}

\newcommand{\msun}{{\rm M}_{\odot}}

\newcommand{\zphot}{z_{\rm phot}}

\shorttitle{UNCOVER NIRSpec $z > 12$}
\shortauthors{Wang et al.}

\begin{document}

\title{UNCOVER: Illuminating the Early Universe --- JWST/NIRSpec Confirmation of $z > 12$ Galaxies}

\correspondingauthor{Bingjie Wang}
\email{bwang@psu.edu}

\author[0000-0001-9269-5046]{Bingjie Wang (\begin{CJK*}{UTF8}{gbsn}王冰洁\ignorespacesafterend\end{CJK*})}
\affiliation{Department of Astronomy \& Astrophysics, The Pennsylvania State University, University Park, PA 16802, USA}
\affiliation{Institute for Computational \& Data Sciences, The Pennsylvania State University, University Park, PA 16802, USA}
\affiliation{Institute for Gravitation and the Cosmos, The Pennsylvania State University, University Park, PA 16802, USA}

\author[0000-0001-7201-5066]{Seiji Fujimoto}\altaffiliation{Hubble Fellow}
\affiliation{
Department of Astronomy, The University of Texas at Austin, Austin, TX 78712, USA
}

\author[0000-0002-2057-5376]{Ivo Labb\'e}
\affiliation{Centre for Astrophysics and Supercomputing, Swinburne University of Technology, Melbourne, VIC 3122, Australia}

\author[0000-0001-6278-032X]{Lukas J. Furtak}
\affiliation{Department of Physics, Ben-Gurion University of the Negev, P.O. Box 653, Be’er-Sheva 84105, Israel}

\author[0000-0001-8367-6265]{Tim B. Miller}
\affiliation{Department of Astronomy, Yale University, New Haven, CT 06511, USA}
\affiliation{Center for Interdisciplinary Exploration and Research in Astrophysics (CIERA) and Department of Physics \& Astronomy, Northwestern University, IL 60201, USA}

\author[0000-0003-4075-7393]{David J. Setton}
\affiliation{Department of Physics \& Astronomy and PITT PACC, University of Pittsburgh, Pittsburgh, PA 15260, USA}

\author[0000-0002-0350-4488]{Adi Zitrin}
\affiliation{Department of Physics, Ben-Gurion University of the Negev, P.O. Box 653, Be’er-Sheva 84105, Israel}

\author[0000-0002-7570-0824]{Hakim Atek}
\affiliation{Institut d'Astrophysique de Paris, CNRS, Sorbonne Universit\'e, 98bis Boulevard Arago, F-75014, Paris, France}

\author[0000-0001-5063-8254]{Rachel Bezanson}
\affiliation{Department of Physics \& Astronomy and PITT PACC, University of Pittsburgh, Pittsburgh, PA 15260, USA}

\author[0000-0003-2680-005X]{Gabriel Brammer}
\affiliation{Cosmic Dawn Center (DAWN), Niels Bohr Institute, University of Copenhagen, Jagtvej 128, K{\o}benhavn N, DK-2200, Denmark}

\author[0000-0001-6755-1315]{Joel Leja}
\affiliation{Department of Astronomy \& Astrophysics, The Pennsylvania State University, University Park, PA 16802, USA}
\affiliation{Institute for Computational \& Data Sciences, The Pennsylvania State University, University Park, PA 16802, USA}
\affiliation{Institute for Gravitation and the Cosmos, The Pennsylvania State University, University Park, PA 16802, USA}

\author[0000-0001-5851-6649]{Pascal A. Oesch}
\affiliation{Department of Astronomy, University of Geneva, Chemin Pegasi 51, 1290 Versoix, Switzerland}
\affiliation{Cosmic Dawn Center (DAWN), Niels Bohr Institute, University of Copenhagen, Jagtvej 128, K{\o}benhavn N, DK-2200, Denmark}

\author[0000-0002-0108-4176]{Sedona H. Price}
\affiliation{Department of Physics \& Astronomy and PITT PACC, University of Pittsburgh, Pittsburgh, PA 15260, USA}

\author[0009-0009-9795-6167]{Iryna Chemerynska}
\affiliation{Institut d'Astrophysique de Paris, CNRS, Sorbonne Universit\'e, 98bis Boulevard Arago, 75014, Paris, France}

\author[0000-0002-7031-2865]{Sam E. Cutler}
\affiliation{Department of Astronomy, University of Massachusetts, Amherst, MA 01003, USA}

\author[0000-0001-8460-1564]{Pratika Dayal}
\affiliation{Kapteyn Astronomical Institute, University of Groningen, 9700 AV Groningen, The Netherlands}

\author[0000-0002-8282-9888]{Pieter van Dokkum}
\affiliation{Department of Astronomy, Yale University, New Haven, CT 06511, USA}

\author[0000-0003-4700-663X]{Andy D. Goulding}
\affiliation{Department of Astrophysical Sciences, Princeton University, Princeton, NJ 08544, USA}

\author[0000-0002-5612-3427]{Jenny E. Greene}
\affiliation{Department of Astrophysical Sciences, Princeton University, Princeton, NJ 08544, USA}

\author[0000-0001-7440-8832]{Y. Fudamoto}
\affiliation{Waseda Research Institute for Science and Engineering, Faculty of Science and Engineering, Waseda University, 3-4-1 Okubo, Shinjuku, Tokyo 169-8555, Japan}
\affiliation{National Astronomical Observatory of Japan, 2-21-1, Osawa, Mitaka, Tokyo, Japan}

\author[0000-0002-3475-7648]{Gourav Khullar}
\affiliation{Department of Physics \& Astronomy and PITT PACC, University of Pittsburgh, Pittsburgh, PA 15260, USA}

\author[0000-0002-5588-9156]{Vasily Kokorev}
\affiliation{Kapteyn Astronomical Institute, University of Groningen, 9700 AV Groningen, The Netherlands}

\author[0000-0001-9002-3502]{Danilo Marchesini}
\affiliation{Department of Physics \& Astronomy, Tufts University, MA 02155, USA}

\author[0000-0002-9651-5716]{Richard Pan}
\affiliation{Department of Physics \& Astronomy, Tufts University, MA 02155, USA}
 
\author[0000-0003-1614-196X]{John R. Weaver}
\affiliation{Department of Astronomy, University of Massachusetts, Amherst, MA 01003, USA}

\author[0000-0001-7160-3632]{Katherine E. Whitaker}
\affiliation{Department of Astronomy, University of Massachusetts, Amherst, MA 01003, USA}
\affiliation{Cosmic Dawn Center (DAWN), Niels Bohr Institute, University of Copenhagen, Jagtvej 128, K{\o}benhavn N, DK-2200, Denmark}

\author[0000-0003-2919-7495]{Christina C. Williams}
\affiliation{NSF's National Optical-Infrared Astronomy Research Laboratory, Tucson, AZ 85719, USA}
\affiliation{Steward Observatory, University of Arizona, Tucson, AZ 85721, USA}

\begin{abstract}

Observations of high-redshift galaxies provide a critical direct test to the theories of early galaxy formation, yet to date, only three have been spectroscopically confirmed at $z>12$. Due to strong gravitational lensing over a wide area, the galaxy cluster field A2744 is ideal for searching for the earliest galaxies. Here we present JWST/NIRSpec observations of two galaxies: a robust detection at $z_{\rm spec} = 12.393^{+0.004}_{-0.001}$, and a plausible candidate at $z_{\rm spec} = 13.079^{+0.013}_{-0.001}$. The galaxies are discovered in JWST/NIRCam imaging and their distances are inferred with JWST/NIRSpec spectroscopy, all from the JWST Cycle 1 UNCOVER Treasury survey. Detailed stellar population modeling using JWST NIRCam and NIRSpec data corroborates the primeval characteristics of these galaxies: low mass ($\sim 10^8~{\rm M_\odot}$), young, rapidly-assembling, metal-poor, and star-forming. Interestingly, both galaxies are spatially resolved, having lensing-corrected rest-UV effective radii on the order of 300--400 pc, which are notably larger than other spectroscopically confirmed systems at similar redshifts. The observed dynamic range of $z \gtrsim 10$ sizes spans over 1 order of magnitude, implying a significant scatter in the size--mass relation at early times. Deep into the epoch of reionization, these discoveries elucidate the emergence of the first galaxies.

\end{abstract}


\section{Introduction}

The cosmic dawn, marked by the emergence of the first stars that formed out of the pristine primordial gas, heralded the most recent global phase transition of the universe: known as the epoch of reionization, it is a period where the first luminous sources ionized the neutral hydrogen that once pervaded the entire universe. While there exist observational constraints coming from the Thomson scattering optical depth inferred from the cosmic microwave background \citep{Hinshaw2013,2020A&A...641A...6P} and quasar absorption lines (\citealt{Fan2006,McQuinn2016}, and references therein) via the Gunn-Peterson effect \citep{Gunn1965}, the history and the drivers of reionization remain largely unknown (see \citealt{Robertson2022:araa} for a recent review). Finding the earliest galaxies constitutes a key piece in constructing a coherent narrative of the cosmic history. Their properties would provide crucial insights into some of the longstanding questions of extragalactic astronomy: the nature of population III stars, the coalescence of the first cosmic structures, and the seeds of supermassive black holes (see \citealt{Bromm2011:araa,Inayoshi2020:araa,Klessen2023:araa} for theoretical overviews).

The past decade has seen substantial advancements in the observations of galaxies at $6<z<10$, offering evidence for early star formation activities and their contribution to reionization \citep{Madau2014:araa,Stark2016:araa}. This progress has mainly been driven by the near-infrared (IR) sensitivity of the Wide Field Camera 3 (WFC3) onboard the Hubble Space Telescope (HST). The most distant known galaxy until recently was GN-z11 with a grism redshift of $z \sim 11$ \citep{Oesch2016}. The redshift frontier is now pushed farther back owing to the launch of the James Webb Space Telescope (JWST), whose NIRCam instrument extends our high-resolution IR coverage to $\sim 1-5$ $\mu$m and also provides higher sensitivity than HST \citep{Rigby2023}. In addition to unambiguously establishing the spectroscopic redshift of GN-z11 at $z = 10.60$ \citep{Bunker2023}, several $z \gtrsim 10$ galaxies have been confirmed via NIRSpec \citep{2023ApJ...951L..22A,2023arXiv230315431A,Roberts-Borsani2023,Fujimoto2023,2023Sci...380..416W} and two at $z>12$, the highest redshift among which is at $z=13.20^{+0.04}_{-0.07}$ \citep{Curtis-lake2023,Robertson2023}. The only other confirmed NIRCam-selected galaxy at $z \sim 12$ is based on an ALMA detection of [O~\textsc{iii}] emission \citep{Bakx2023}. These discoveries provide some of the first direct insights into the astrophysical processes governing galaxy growth near the cosmic dawn.

Aided by strong gravitational lensing, the JWST Treasury Cycle 1 UNCOVER survey \citep{Bezanson2022} turns the field containing the galaxy cluster Abell~2744 at $z=0.308$ into the deepest view of our universe to date---the intrinsic depths of $\gtrsim31-32$ AB magnitudes make it ideal for carrying out searches for the first galaxies. In this Letter, we present two high-redshift galaxies discovered with UNCOVER spectroscopy. \twelve\ has a secure \zspectw\ due to an unambiguous Lyman break; \thirteen\ is tentatively detected at \zspecth. The lower signal-to-noise ratio (S/N) of the latter leads us to refrain from a highly confident $z>13$ interpretation, although the probability density distribution of redshift, $p(z)$, sharply peaks at this high redshift.

The structure of this Letter is as follows. Section~\ref{sec:data} provides an overview of the data, including imaging, candidate selection, and spectroscopy. Section~\ref{sec:data_analyses} details the measurement of the spectroscopic redshift and of the  morphology in the imaging. Section~\ref{sec:sed_fitting} explains the assumptions and components involved in fitting the spectral energy distribution (SED). Section~\ref{sec:res} presents the results. Section~\ref{sec:conclu} concludes with a summary of the key findings.

Where applicable, we adopt the best-fit cosmological parameters from the WMAP 9 yr results: $H_{0}=69.32$ ${\rm km \,s^{-1} \,Mpc^{-1}}$, $\Omega_{M}=0.2865$, and $\Omega_{\Lambda}=0.7135$ \citep{Hinshaw2013} and a Chabrier initial mass function (IMF; \citealt{Chabrier2003}) over 0.1-100 $\msun$.
Unless otherwise mentioned, we report the median of the posterior, and 1$\sigma$ error bars are the 16th and 84th percentiles.

\section{Data\label{sec:data}}

\subsection{Imaging}

The reduction of all publicly available HST and JWST imaging over Abell~2744 used herein is outlined in \citet{Bezanson2022}. This comprises 15 broad- and medium-band filters from JWST/NIRCam, HST/ACS, and HST/WFC3, covering the wavelength range of $\sim 0.4-5~\mu$m. The JWST/NIRCam observations originate from three programs: UNCOVER (PIs Labbe \& Bezanson, JWST-GO-2561; \citealt{Bezanson2022}), the Early Release Science program GLASS (PI: Treu, JWST-ERS-1324; \citealt{Treu2022}), and a Director's Discretionary program (PI: Chen, JWST-DD-2756), together providing 8 filters: F090W, F115W, F150W, F200W, F277W, F356W, F410M, and F444W. The HST data, obtained from the public archive, include HST-GO-11689 (PI: Dupke), HST-GO-13386 (PI: Rodney), HST-DD-13495 (PI: Lotz; \citealt{Lotz2017}), and HST-GO-15117 (PI: Steinhardt; \citealt{Steinhardt2020}). These additional observations cover a wavelength range of $\sim 0.4-1.6 \mu$m in the observed frame, using 7 filters: F435W, F606W, F814W, F105W, F125W, F140W, and F160W.

Photometric catalogs are constructed for $\sim$~50,000 objects identified from a noise-equalized detection image combining F277W, F356W, and F444W imaging, with aperture-corrected fluxes extracted with the \texttt{python} implementation of Source Extractor \citep{1996A&AS..117..393B,Barbary2016} as described in \citet{Weaver2023}. We use the latest photometry herein (internal release v3.0.0)\footnote{Since making the preprints available in early 2023, both the data reduction and photometric pipelines have improved significantly, details of which will be presented in the published versions of \citet{Bezanson2022} and \citet{Weaver2023}.}. The photometric measurements for both sources are extracted within an aperture size of 0.32\arcsec{} on images convolved to a uniform point spread function.

For the lens model, we use an updated version of the \citet{Furtak2023} analytic model of Abell~2744. This version\footnote{The \texttt{v1.1} deflection maps are publicly available on the UNCOVER website \url{https://jwst-uncover.github.io/DR1.html\#LensingMaps}.} includes one additional multiple image system in the northern sub-structure, and more importantly, an additional spectroscopic redshift in the north-western sub-structure from new VLT/MUSE observations of the cluster \citep{Bergamini2023b}. These improvements bring the lens plane image reproduction root-mean-square (RMS) of the model down to $\Delta_{\mathrm{RMS}}=0.51\arcsec$.

\subsection{Candidates}

The sample consists of galaxies initially identified in HST + JWST imaging. \twelve\ is originally identified in \citet{Atek2023:uncover} (ID 42329), with a photometric redshift at $z_\mathrm{phot}=11.83^{+1.05}_{-7.93}$. The selection procedure combines dropout information from the color-color diagram, and photometric redshifts using the \eazy\ and \beagle\ software. In particular, a strong Lyman break ($>$~1.5 AB mag) is required to meet the selection threshold. A secondary set of criteria addresses the quality of the photometric redshifts: here, $\chi^{2} < 30$ combined with good agreement (within uncertainties) between \eazy\ and \beagle\ solutions is enforced. \twelve\ is assigned a quality flag of 3 mainly due to disagreements between the photometric redshift solutions (see \citealt{Atek2023:uncover} for details).

\thirteen\ is selected based on two main criteria. The first is the reliability of the photometry. This includes \verb|use_phot|=1 (see \citealt{Weaver2023} for details on the flag), S/N $>$~7 in the long-wavelength filters bands, and having data in at least 6 NIRCam filter bands. Second, we require a robust high-redshift solution from the photometry, with a redshift posterior median at $\zphot > 11$ and the 16th percentile of the redshift posterior $> 6$ (i.e., a low probability of low-redshift interlopers). There is a final visual inspection for basic quality control before the allocation of NIRSpec slits. In addition, the \prospector-$\beta$ model \citep{Wang2023:pbeta} within \prospector\ \citep{Johnson2021} finds both candidates to have $\zphot > 11$ \citep{Wang2023:uncover}.

In total, there are eight $z>11$ candidates followed up with spectroscopy, further details of which can be found in Appendix~\ref{app:sample}. Low- and high-redshift galaxies possibly sharing similar photometric colors is a known challenge, and is exacerbated by the new redshift frontier in the era of JWST \citep{Dunlop2007,Naidu2022dust,2023ApJ...951L..22A,McKinney2023,Zavala2023}. The redshift calibration for the UNCOVER survey will be presented in a future paper.

\begin{figure*} 
 \gridline{
 \fig{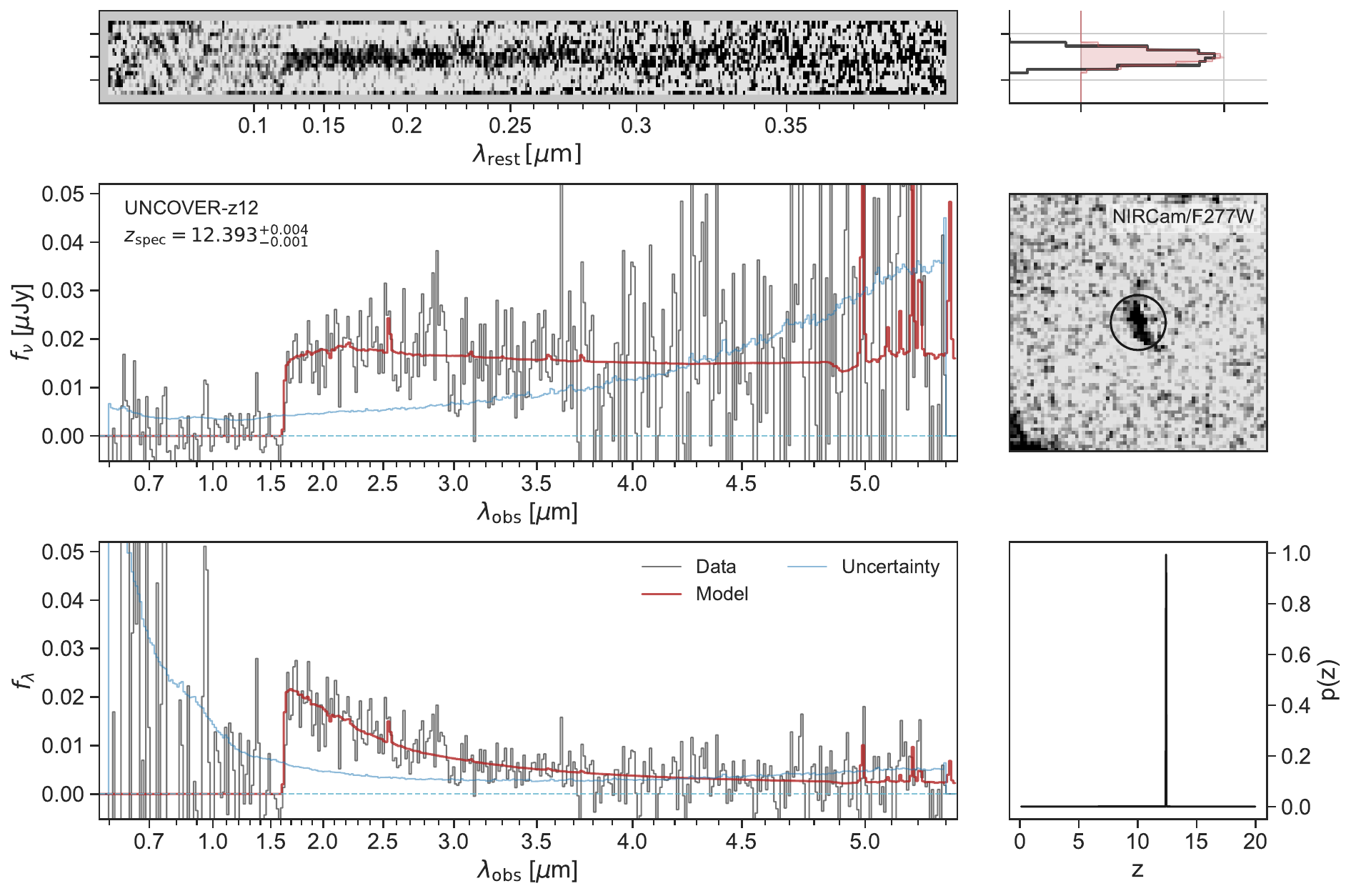}{0.8\textwidth}{(a)}
 }
 \gridline{
 \fig{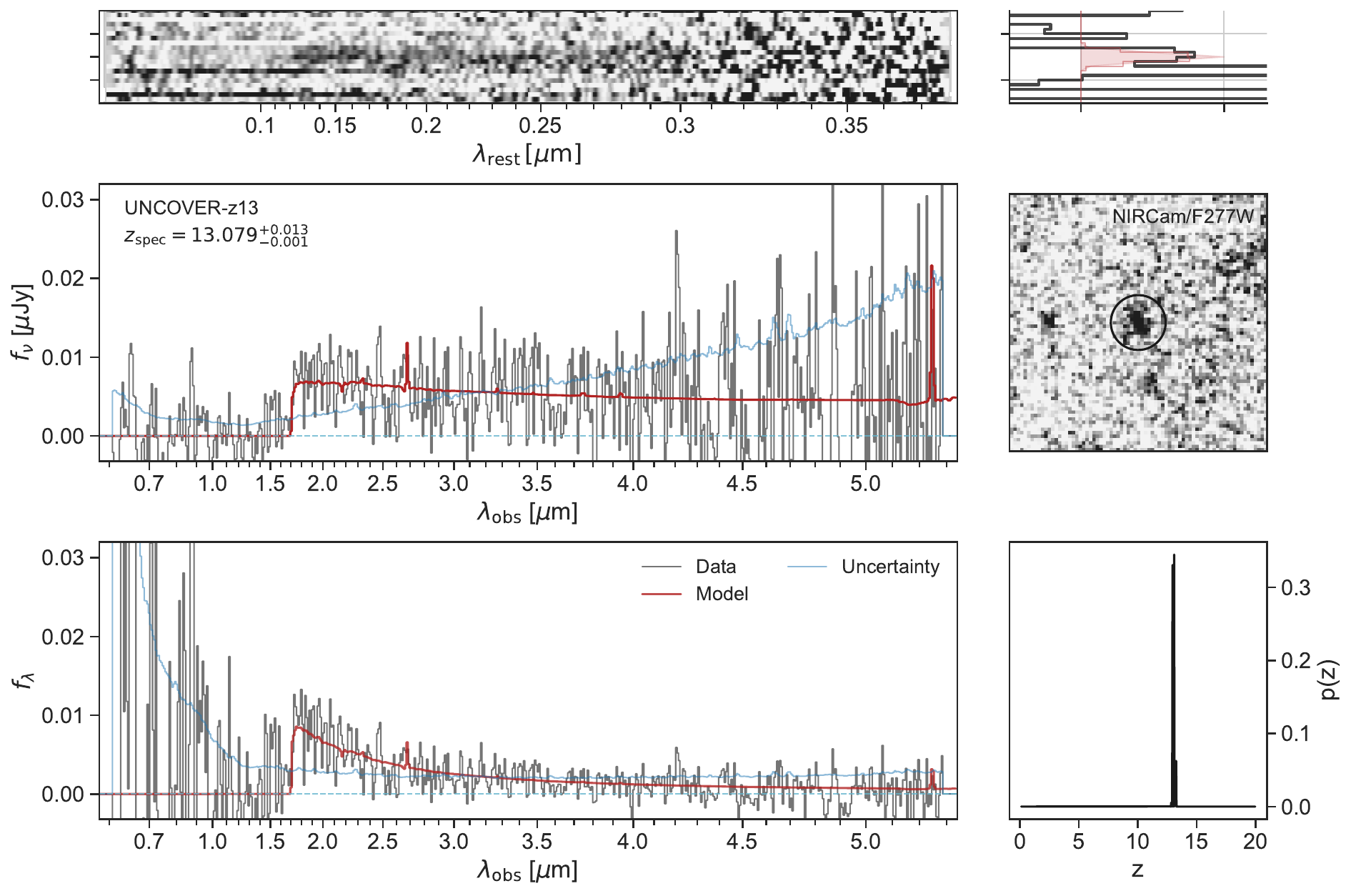}{0.8\textwidth}{(b)}
 } 
\caption{NIRSpec/Prism observations of the two galaxies presented in this Letter. (a) The first row displays the 2D spectrum of \twelve, with the rest-frame wavelength shown along the $x$-axis; the histogram indicates the extracted region. Second row, left panel, shows the 1D spectrum in $f_\nu$ as a function of the observed wavelength. Data are plotted in gray, whereas the best-fit \eazy\ model is plotted in red. Also plotted are the uncertainties in blue. The cyan horizontal line is at $y=0$ to guide the eye. Second row, right panel, includes the cutout in the F277W band, with the circle indicating the aperture size, 0.32\arcsec{}, in which photometry is extracted. Third row, left panel, shows the 1D spectrum again, but in $f_\lambda$ as a complementary visualization. Third row, right panel, plots the probability density over the redshift range of the search for a minimum $\chi^2$. A strong Lyman break is clearly seen for \twelve. (b) Same as the above figure set, but for \thirteen. We err on the conservative side and consider its redshift solution to be less well-determined due to the weaker break, even though $p(z)$ sharply peaks at $z > 12$.}
\label{fig:data}
\end{figure*}

\subsection{NIRSpec/Prism Spectroscopy}

The JWST/NIRSpec low-resolution Prism spectra presented in this paper were collected between 2023 July 31 and August 2 as part of the second phase of the UNCOVER Treasury survey \citep{Bezanson2022}. This Letter is based on internal v0.3 reductions. The full spectroscopic data reduction and extraction will be presented in Price et al. in prep. We summarize briefly as follows. These NIRSpec/MSA observations are separated into 7 semi-overlapping footprints with some repeated targets, yielding on-source integration times of between 2.7 and 17.4 hr. Overall integration times for the current targets are 4.4 hr (38766, MSA4) and 7.4 hr (13077, MSA5 and MSA7). All sources were assigned three-slitlets and observations were conducted with a \verb|2-POINT-WITH-NIRCam-SIZE2| dither pattern.

Stage 2 data products were downloaded from MAST reduced with \texttt{msaexp} (v0.6.10; \citealt{Brammer2022}). \texttt{msaexp} masks artifacts (including snowballs), and corrects for 1/$f$ noise. Individual slits are identified and WCS solutions are applied, data are flat fielded and background subtracted (from vertically shifted spectra). The 2D spectra are stacked and drizzled together and then optimally extracted. 
We also perform an additional manual extraction on \thirteen{} as a complementary quality control. 
Both 1D and 2D spectra are included in Figure~\ref{fig:data}.

\section{Data Analyses\label{sec:data_analyses}}

\subsection{Spectroscopic Redshifts \label{subsec:data:zspec}}

Spectroscopic redshifts are determined using \msae, which finds the best-fit redshift by minimizing the $\chi^2$ between a template set and the observed spectrum. We choose the \texttt{blue\_sfhz\_13} template made available as a part of the \eazy\ code package \citep{Brammer2008}. This template set forbids unphysical star formation histories (SFHs) that are older than the age of the universe at the given redshift, making it more suitable for our application. At $z \gtrsim 10$, the strongest feature to secure a redshift solution is the Lyman break. Emission lines are also identified and contribute to the likelihood. The attenuation of the intergalactic medium (IGM) is assumed to follow \citet{Inoue14}, and the damping Lyman-$\alpha$ wing is also considered  \citep[e.g.,][]{2023arXiv230315431A,Curtis-lake2023}.
We allow the search for the minimal $\chi^2$ value over the full range of $0.1 \leq z \leq 20$. Uncertainties are from the 68.2\% range of the integrated $p(z)$. In addition, we characterize the break strength by comparing the mean of the flux in two segments over 150~\AA\ blue- and red-ward of the observed break.

\subsection{Morphology}

Morphology measurements are performed using \texttt{pysersic} \citep{Pasha2023} assuming a Sersic profile for each source and a flat background that is fit simultaneously. Priors for $r_{\rm eff}$, $n_{\rm Sersic}$ and the axis ratio are flat and bounded by 0.5 -- 10 pixels, 0.65 -- 4 and 0.1 -- 1, respectively; priors for the central position and fluxes are Gaussian and based on the measurements from the photometric catalog. The posteriors for all variables are explored using the No U-turn Sampler (NUTS; \citealt{2011arXiv1111.4246H,2019arXiv191211554P}) where two chains are run for 1,000 warm-up iterations followed by 1,000 sampling steps each.

Additionally, we model \twelve\ with two Sersic components. The priors in this case are independent and similar to the above, except for the flux. We assume each source has half of the catalog flux, and the central positions are offset by about two pixels in the northwest (NW) and southeast (SE) direction, respectively. 

Our morphology measurements focus on F277W as this leads to the best constraints for both sources, owing to the higher S/N. It is worth noting that at the redshifts measured, the rest-frame wavelength of the F277W filter is in the far-ultraviolet which is expected to trace recent ($\lesssim$ 20 Myr) unobscured star formation. We also perform Sersic fits in the redder F444W filter, which give similar results for \twelve{}. However, given the faintness of \thirteen{} in the redder bands, the fits for this source yield limited constraints.

\section{SED fitting\label{sec:sed_fitting}}

\subsection{Prospector\label{subsec:prosp}}

The available JWST/NIRCam and HST/ACS, WFC3 photometry are jointly fitted with the full NIRSpec/Prism spectrum with a 18-parameter model within the \prospector\ inference framework \citep{Johnson2021}. Before fitting, all models are convolved with the NIRSpec/Prism instrumental resolution curve~\footnote{\texttt{jwst\_nirspec\_Prism\_disp.fits} provided by the Space Telescope Science Institute}, assuming that the flight performance is 1.3 times better than stated. This accounts for the fact that the spatial extent of the sources in the dispersion direction is narrower than the shutter width, and is consistent with an earlier work where a factor 1/0.7 is introduced for modeling $z>10$ galaxies with \beagle\ \citep{Curtis-lake2023}. As the spectral calibration is at an early stage, we fit for a polynomial calibration vector of order 2 after applying a wavelength-independent calibration to scale the normalization of the spectrum to the photometry.

We follow the same methodology as detailed in the paper presenting the UNCOVER stellar population catalog \citep{Wang2023:uncover}, with the exception that the redshift is restricted to vary in a narrow range around the best-fitting spectroscopic redshifts ($\pm$~0.1) determined by \msae\ based on \eazy\ templates. Here we briefly reiterate the modeling process for completeness. 

All parameters are inferred jointly, using the MIST stellar isochrones \citep{Choi2016,Dotter2016} and MILES \citep{Sanchez-Blazquez2006} stellar spectral library in FSPS \citep{Conroy2010}. The SFH is described by the non-parametric \prospector-$\alpha$ model via mass formed in 7 logarithmically-spaced time bins \citep{Leja2017}. A mass function prior, and a dynamic SFH($M, z$) prior are included to optimize the photometric inference over the wide parameter space covered by deep JWST surveys \citep{Wang2023:pbeta}. Lensing magnification using the publicly-released maps \citep{Furtak2023}, is performed on-the-fly (i.e, $\mu = \mu(z)$) to ensure consistency with the scale-dependent priors. Because of the lack of emission lines observed, we do not turn on the emission-line marginalization. A 5\% error floor is imposed on the photometry to account for systematics. Sampling is performed using the dynamic nested sampler \texttt{dynesty} \citep{Speagle2020}.

\subsection{Bagpipes}

We also perform spectrophotometric modeling using the \texttt{Bagpipes} software package (Bayesian Analysis of Galaxies for Physical Inference and Parameter EStimation; \citealt{Carnall2018,Carnall2019:VANDELS}). 
The following model grid is utilized: \cite{Bruzual:2003} stellar population models, the MILES spectral library \citep{Sanchez-Blazquez2006,Falcon-Barroso2011}, \texttt{Cloudy} nebular emission models \citep{Ferland:2017}, and \citet{Charlot2000} dust model with $0<A_{V}<5$ and $0.3<n<2.5$ as free parameters. The stellar and gas phase metallicity are tied to the same value, and also included as a free parameter. The ionization parameters vary in the ranges of $-2<\mathrm{log}~Z<0.3$ and $-3.5<\mathrm{log}~U<-1.0$. The SFH is parameterized as delayed-$\tau$ (i.e., SFR $\propto \ t e^{-t/ \tau}$) with age and $\tau$ as free parameters ($0.1 \ \mathrm{Gyr}<\mathrm{age} <t_\mathrm{universe}$ and $0.01 \ \mathrm{Gyr}<\tau<5 \ \mathrm{Gyr}$). Redshift likewise varies around the best-fitting spectroscopic redshifts ($\pm$~0.1). 

The same procedure mentioned in Section~\ref{subsec:prosp} regarding the spectral calibration is applied. The sampling is performed via \texttt{PyMultinest} \citep{Buchner:2014,Feroz:2019}, with the default \texttt{Bagpipes} convergence criteria.

\subsection{BEAGLE}

The magnification-corrected spectra are in addition fitted with \beagle\ (BayEsian Analysis of GaLaxy sEds; \citealt{Chevallard2016}), which combines the latest version of the \cite{Bruzual:2003} stellar population models with nebular emission templates by \citet{Gutkin2016} and assumes the \citet{Inoue14} IGM attenuation models. Assuming a delayed-$\tau$ SFH as in our \bagpipes\ fit, but with an additional parameter to independently control the recent SFR (0-10 Myr), and an SMC dust attenuation law \citep{Pei92}, we leave all other \beagle\ parameters free to vary with uniform or log-uniform priors. The free parameters are the stellar mass $\log(M/\mathrm{M}_{\odot})\in[4,10]$, the current (10~Myr) SFR $\log(\psi/\mathrm{M}_{\odot}\,\mathrm{yr}^{-1})\in[-2,4]$, the maximum stellar age $\log(t_{\mathrm{age}}/\mathrm{yr})\in[6,t_{\mathrm{universe}}]$, the star-formation e-folding time $\log(\tau/\mathrm{yr})\in[5.5,9.5]$, the stellar metallicity $\log(Z/\mathrm{Z}_{\odot})\in[-2.2,-0.3]$, the effective \textit{V}-band dust attenuation optical depth $\hat{\tau}_V\in[0,3]$, the effective galaxy-wide ionization parameter $\log~U\in[-4,-1]$, the gas-phase metallicity $\log(Z_{\mathrm{gas}}/\mathrm{Z}_{\odot})\in[-2.2,-0.3]$, and the dust-to-metal mass ratio $\xi_{\mathrm{d}}\in[0.1,0.5]$.

We note that \beagle\ does not calibrate the spectrum to the photometric data during modeling fitting. Given that the absolute flux calibration of NIRSpec is still an ongoing effort in the community, we only consider the \beagle\ results as a sanity check in this work.

\section{Results\label{sec:res}}

The redshifts and inferred stellar population parameters are summarized in Table~\ref{tab:sps}. We discuss the results in more detail in the following sections.

\subsection{Spectroscopic Redshifts}	

The strong break observed in \twelve, seen in Figure~\ref{fig:data}, places it at \zspectw\ in the \msae\ fits. 
The posterior distributions of redshift from both \prospector\ and \bagpipes\ spectrophotometric fits show well-behaved Gaussian shapes centered around the spectroscopic redshift, further supporting the $z>12$ measurement, though here the redshift is confined to within $\pm~0.1$ of the best-fit \msae\ solution by construction. A careful analysis of possible emission lines results in no significant detection. However, comparing the S/N of the mean flux red-ward of the observed break to that blue-ward of the break (3.43 and -0.26, respectively) leads us to conclude that the Lyman break is prominent enough to determine the redshift without ambiguity. A forced Balmer break fit is included in Appendix~\ref{app:balmer}, offering further support for the high-redshift interpretation.

In \thirteen, we report a tentative spectroscopic redshift of \zspecth. It exhibits a break feature at lower S/N, at a level that is analogous to the $z=13.2$ galaxies discovered in JADES \citep{Robertson2023,Curtis-lake2023}. Fits on both of the individual, manually extracted spectra in addition to the co-added spectrum also favor a high-redshift solution. A forced Balmer break fit likewise leads to a higher $\chi^2$. However, while the S/N of the mean flux red-ward of the observed break is similar to the above case, the S/N blue-ward of the break is 1.05, indicating a marginal probability for a Balmer break. Without additional evidence in support of the high redshift interpretation (i.e., significant emission lines), we decide to only make a tentative redshift identification at this time.

\subsection{Morphology}

\begin{figure*} 
 \gridline{
 \fig{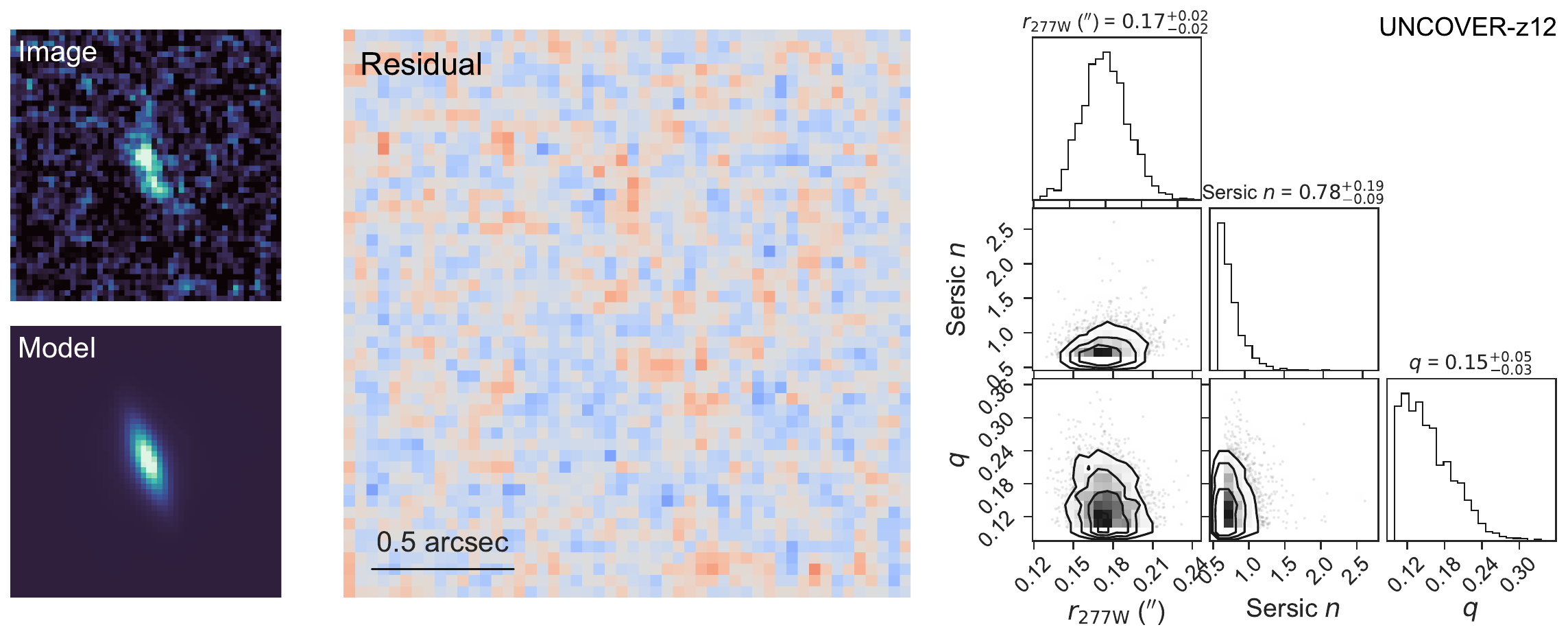}{0.9\textwidth}{(a)}
 }
 \gridline{
 \fig{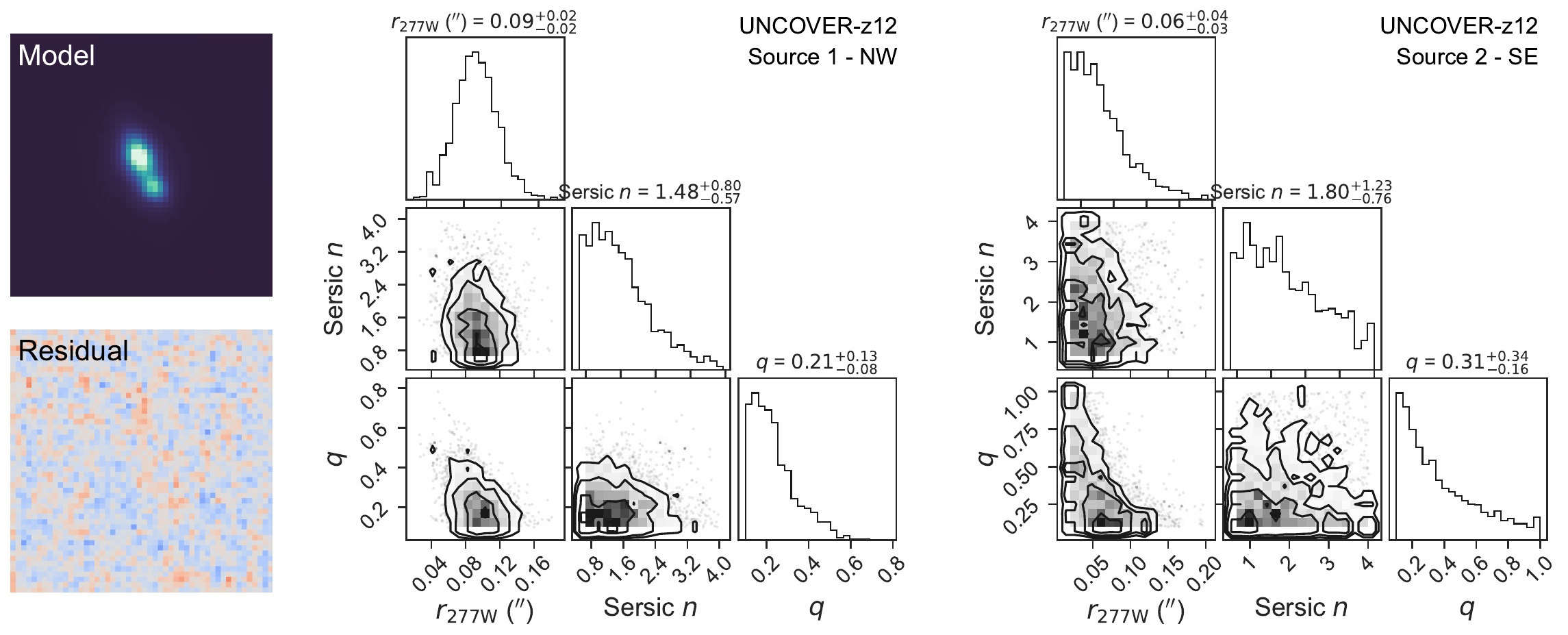}{0.9\textwidth}{(b)}
 }
\gridline{
 \fig{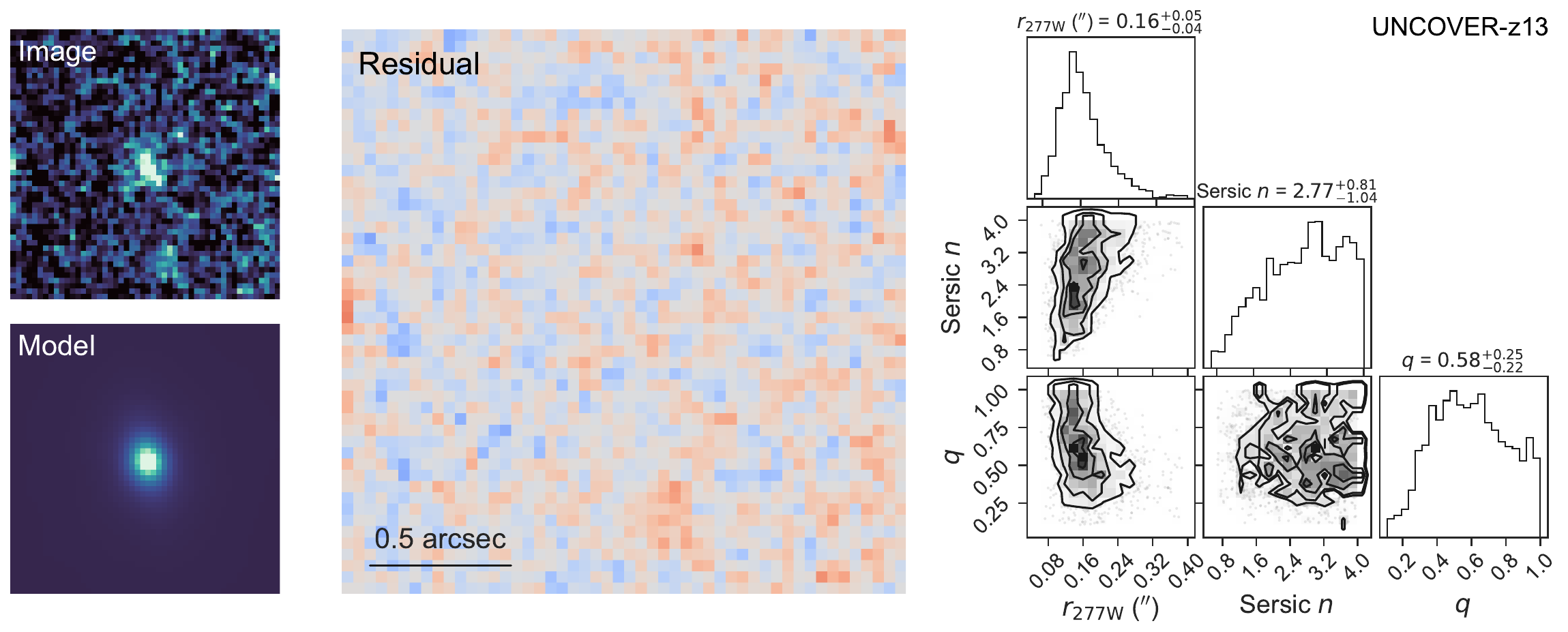}{0.9\textwidth}{(c)}
 }
 \caption{Morphological modeling. (a) \twelve\ is first modeled as a single source. The upper left panel shows its image in F277W, tracing recent unobscured star formation, whereas the lower left panel shows the single-component model fit. The model residual, scaled by the RMS, is shown in the middle. A corner plot illustrating the posterior distributions is also included. 
 (b) \twelve\ is also modeled as two components, and the results are presented in a similar manner.
 (c) Same as the first figure set, but for \thirteen.}
\label{fig:morph}
\end{figure*}

Both \twelve{} and \thirteen{} are clearly resolved with measured half-light radii of $4.35 ^{+0.40} _{-0.39}$ pixels and $3.95 ^{+1.37} _{-0.94}$ pixels respectively, with a pixel scale of 0.04\arcsec per pixel. The results are presented in Figure~\ref{fig:morph}, and we elaborate on each galaxy below.

\twelve\ shows a disk-like morphology with a measured Sersic index of  $0.78 ^{+0.19} _{-0.09}$ and an axis ratio of $0.15 ^{+0.05} _{-0.03}$ --- this is very elongated and approaches the lower bound of the prior of 0.1. Shear boosts the semi-major axis by $\mu_{\rm t} = 1.52 ^{+0.02} _{-0.03} $, and has roughly no effect on the semi-minor axis, based on the lens model \citep{Furtak2023}. This suggests an intrinsic axis ratio of $\sim 0.23$. There appears to be two main clumps in the F277W image of \twelve{}, which can contribute to the extended size. In F444W, the galaxies show a much smoother distribution with a Sersic fit that leads to similar constraints. There are no obvious systematic differences in the residuals from the Sersic fit to the F277W image. The disturbed morphology can be driven by clumpy ongoing star formation, as it is more prevalent at shorter wavelengths.

However, an ongoing merger could also explain the clumpy morphology, we thus model \twelve{} with two Sersic components simultaneously as shown in Figure~\ref{fig:morph}-b. We find the brighter Source-1 in the NW contains about 2/3 of the total flux of the system, which remains resolved but with a half-light radius roughly half of that implied by the single Sersic fit. Source-1 remains elongated along a similar axis and still shows disk morphology with a Sersic index consistent with one and a lensing-corrected axis ratio of 0.31. Source-2 in the SE is much fainter, resulting in a $1\sigma$ upper limit on the size of about 1.5 pixels, or 0.06$\arcsec$. Its Sersic index and axis ratio are unconstrained. We note that the resulting $\chi ^2$ values are nearly identical when comparing the single Sersic fit to the two Sersic fits, with the latter being lower by only 2\%. With both cases being physically plausible, and neither being preferred by the data, we consider both as limiting cases for the remainder of the study.

As for \thirteen, it has a Sersic index of $2.77 ^{+0.81}_{-1.04}$ with a poorly constrained axis ratio of $0.58 ^{+0.25}_{-0.21}$. Similarly, shear boosts the semi-major axis by $\sim$~1.9, and the semi-minor axis by $\sim$~1.5.

\subsection{Stellar Populations}

\begin{figure*} 
 \gridline{
 \fig{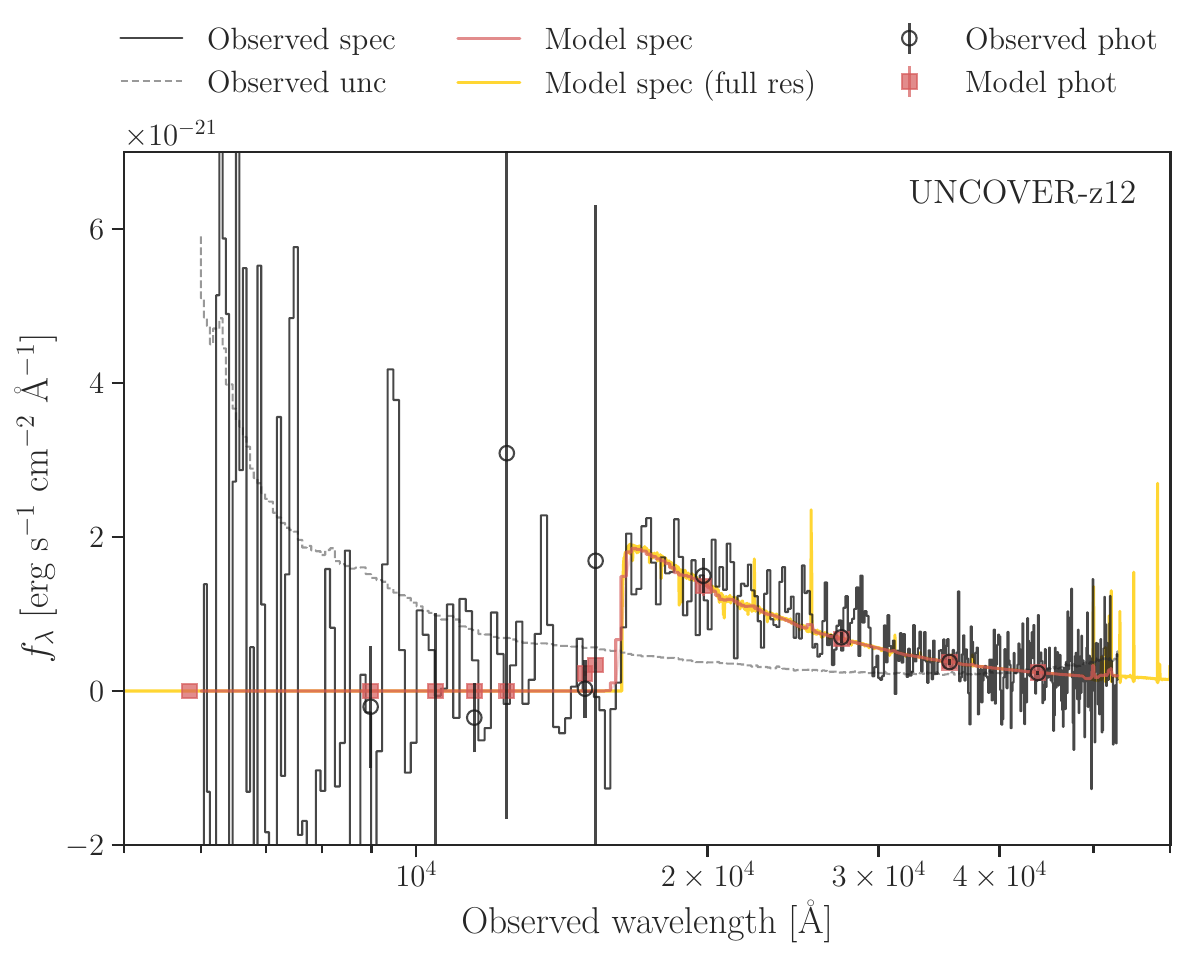}{0.5\textwidth}{(a)}
 \fig{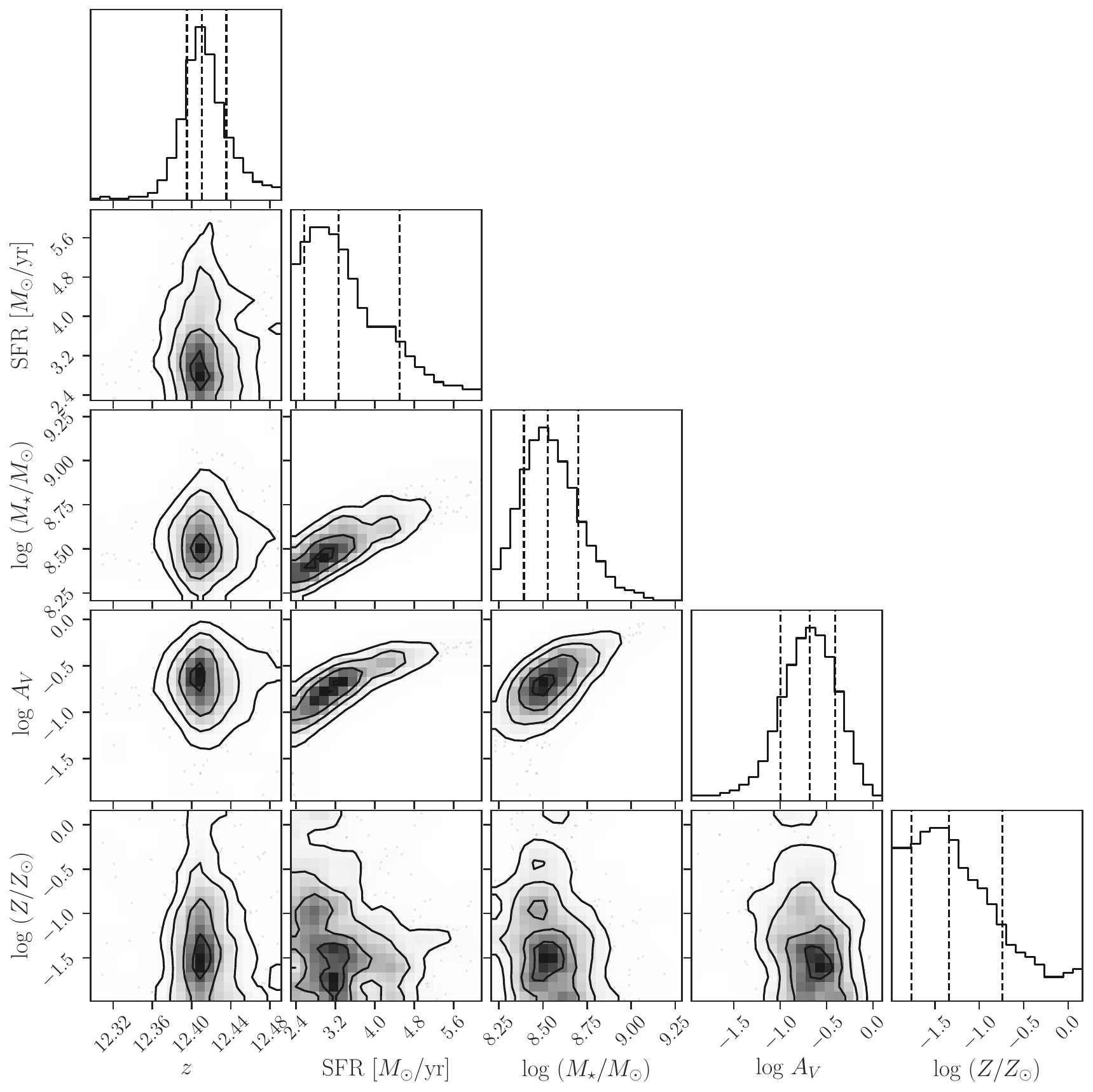}{0.45\textwidth}{(b)}
 }
 \gridline{
 \fig{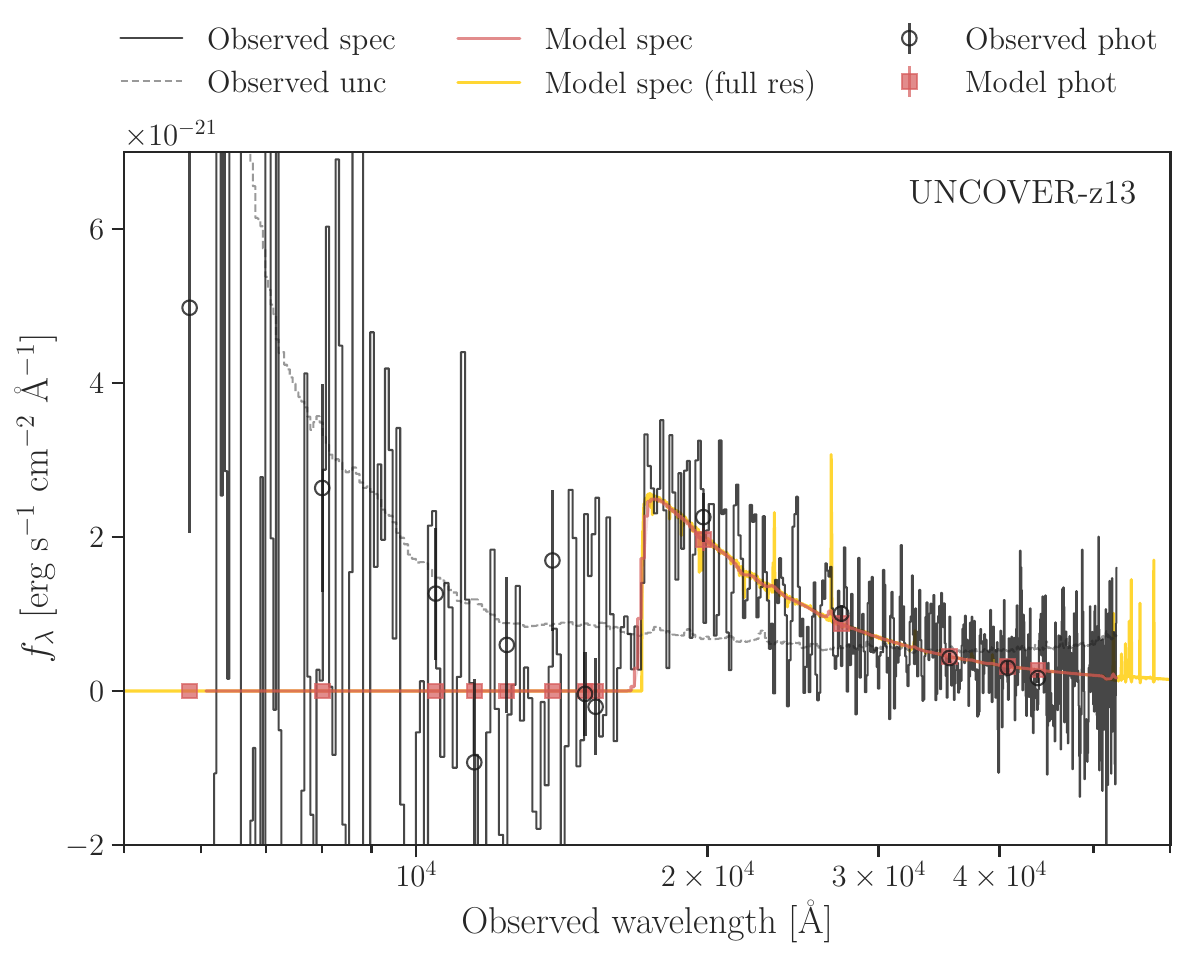}{0.5\textwidth}{(c)}
 \fig{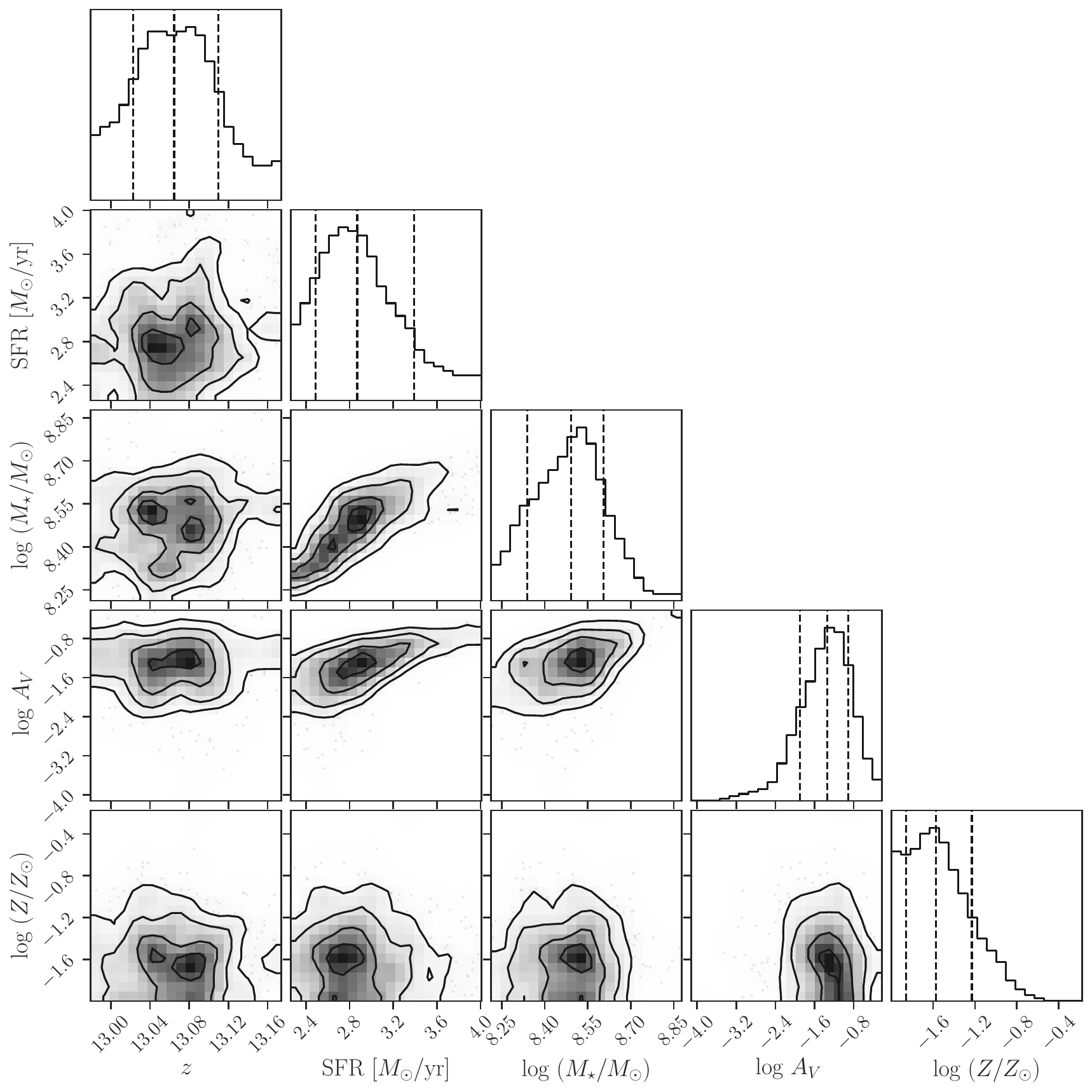}{0.45\textwidth}{(d)}
 }
 \caption{Stellar population modeling with \bagpipes. (a) The median model spectrum and photometry are shown in red, whereas the observations are shown in black. The model spectrum at the full resolution is additionally shown in gold, highlighting the predicted emission features that are washed out by the instrumental resolution. Both model spectra are scaled by a polynomial calibration vector to match the photometry. (b) The corner plot illustrates posterior distributions for a few key parameters; scale-dependent parameters are magnification-corrected. (c)--(d) Same as the above figure set, but for \thirteen. We note that the photometric data points at short wavelengths are consistent with zero flux at 1.5$\sigma$ confidence.}
\label{fig:bagpipes}
\end{figure*}

The different galaxy SED-fitting codes in general agree on the inferred properties for both objects, details of which are listed in Table~\ref{tab:sps}. The best-fit model spectra from \bagpipes, alongside posterior distributions of key parameters, are shown in Figure~\ref{fig:bagpipes}. All model parameters are corrected for the magnification factor, $\mu$. Both galaxies are moderately magnified with $\mu \sim 2$, meaning that the observed uncertainties of the photometric and spectral fluxes dominate. We find the galaxies of this work have low stellar masses of $\sim 10^8 \msun$, similar to the others discovered at these redshifts (e.g., \citealt{Robertson2023}). They are young and metal-poor; the metallicity constraints, while highly uncertain, are slightly higher than the similar spectroscopically confirmed $z>10$ systems presented in \cite{Curtis-lake2023}. The solutions also favor fairly low dust content ($A_V\sim0.1$) and high specific SFRs ($\mathrm{log~sSFR}/\mathrm{yr}\sim-8$), comparable to other high-redshift systems.

In brief, these characteristics are in alignment with our expectations of early galaxies. For example, for $z \sim 12.2$ sources with $M_* \sim 10^{7.9-8.4} \msun$, semianalytic galaxy formation models DELPHI \citep{2014MNRAS.445.2545D,2023arXiv230501681M}, base-lined against data at $z \sim 5-10$, would predict SFRs between 0.8 and 3 $\msun \rm{yr^{-1}}$ and associated gas-phase metallicity values between -1.8 and -0.8. At $z \sim 13$, for $M_* \sim 10^{7.6-8.1} \msun$, this model would predict SFR between 0.5 and 2.6 $\msun \rm{yr^-1}$ and associated gas-phase metallicities between -1.8 and -0.9, in log$_{10}$ units relative to solar metallicity. Within uncertainties, these values are in good agreement with those inferred from our SED fitting, suggesting that these galaxies lie on or close to the values expected for main-sequence systems at these early epochs.

\begin{deluxetable}{lccc}
\tablecaption{Ancillary Parameters\label{tab:sps}}
\tablehead{
\colhead{} & \colhead{\twelve{}\tablenotemark{\scriptsize{a}}} & \colhead{\thirteen{}\tablenotemark{\scriptsize{b}}} }
\startdata
RA [deg] & 3.51356 & 3.57087 \\
Dec [deg] & -30.35680 & -30.40158 \\
$z_{\rm spec}$ & $12.393_{-0.001}^{+0.004}$ & $13.079_{-0.001}^{+0.013}$ \\
$M_{\rm F444W}$ &  -18.9$\pm$0.1  & -18.3$\pm$0.6 \\
$M_{\rm F200W}$ &  -19.2$\pm$0.5  & -19.4$\pm$1.8 \\
$\mu$ & $1.520_{-0.039}^{+0.015}$ & $2.278_{-0.005}^{+0.084}$ \\ 
$\mu_{\rm t}$ & $1.511_{-0.029}^{+0.018}$ & $1.881_{-0.012}^{+0.044}$ \\ 
Sersic $n$ & $2.77_{-1.04}^{+0.81}$ & $0.78_{-0.09}^{+0.19}$ \\
\hline 
\prospector \\
log~$M/\msun$  & $8.63^{+0.20}_{-0.22}$ & $7.92^{+0.23}_{-0.16}$ \\
Age [yr] & $7.76^{+0.34}_{-0.19}$ & $7.65^{+0.22}_{-0.29}$ \\
SFR [$\msun\,{\rm yr}^{-1}$] & $1.40^{+2.81}_{-1.16}$ & $1.95^{+0.41}_{-0.39}$ \\
log~sSFR [${\rm yr}^{-1}$] & $-8.69^{+0.68}_{-0.93}$ & $-7.76^{+0.23}_{-0.33}$ \\
log~$Z/{\rm Z}_\odot$ & $-1.40^{+0.67}_{-0.39}$ & $-1.19^{+0.87}_{-0.53}$ \\
$\hat\tau_{\rm dust,2}$ & $0.17^{+0.34}_{-0.13}$ & $0.02^{+0.04}_{-0.02}$ \\
\hline 
\texttt{Bagpipes}\\
log~$M/\msun$ & $8.35^{+0.18}_{-0.14}$ & $8.13^{+0.11}_{-0.15}$ \\
Age [yr] & $7.79^{+0.18}_{-0.15}$ & $7.82^{+0.14}_{-0.17}$ \\
SFR [$\msun\,{\rm yr}^{-1}$] & $2.15^{+0.81}_{-0.46}$ & $1.28^{+0.27}_{-0.18}$ \\
log~sSFR [${\rm yr}^{-1}$] & $-8.00^{+0.07}_{-0.11}$ & $-8.01^{+0.08}_{-0.10}$ \\
log~$Z/{\rm Z}_\odot$ & $-1.34^{+0.60}_{-0.42}$ & $-1.57^{+0.35}_{-0.28}$ \\
$\hat\tau_{\rm v}$ & $0.19^{+0.17}_{-0.10}$ & $0.04_{-0.03}^{+0.08}$ \\
\enddata
\tablenotetext{a}{MSA ID 38766.}
\tablenotetext{b}{MSA ID 13077.}
\end{deluxetable}

\section{Discussion and Conclusions\label{sec:conclu}}

This Letter presents spectroscopic observations of two galaxies using NIRSpec/Prism onboard JWST. We summarize the main findings as follows. 

\twelve\ is located securely at \zspectw\ due to an unambiguous Lyman break, placing it as the fourth most distant, spectroscopically confirmed galaxy known to date; \thirteen\ is tentatively detected at \zspecth. We err on the conservative side and do not firmly rule out a lower-redshift solution for \thirteen\ for two reasons: first, the break is weaker, and thus possibly instead a Balmer break; second, the emission features seen in the spectrum are consistent with noise, providing no additional evidence in support of a $z\sim13$ interpretation.

Stellar population modeling is performed using independent codes, all constrained by the same HST+JWST photometric and JWST/NIRSpec spectroscopic observations. The inferred galaxy properties are consistent with expectations of early systems: low mass ($\sim 10^8 \msun$), young, rapidly-assembling, metal-poor, and star-forming.

As both galaxies are clearly spatially resolved, we are able to make reliable morphological measurements. The shear-corrected effective radii are $426^{+40}_{-42}$ pc and $309^{+110}_{-74}$ pc for \twelve\ and \thirteen, respectively. The resulting SFR surface densities are $\sim 2.45 \pm 0.85$ and $1.68\pm 1.00 ~\msun {\rm yr}^{-1} {\rm kpc}^{-2}$, which are less extreme than the 15 -- 180 $\msun {\rm yr}^{-1} {\rm kpc}^{-2}$ range reported in \citet{Robertson2023}. In addition, \twelve\ exhibits a disturbed morphology, indicating clumpy star formation \citep{Marshall2022,Chen2023,Trussler2023}. We thus consider an alternative merger scenario, which is likely in the context of hierarchical structure formation \citep{White1991}. This interaction would trigger star formation, manifesting as burstiness in the inferred SFH. With that being said, the observations place limited constraints on the SFH, offering no additional evidence for or against the merger hypothesis. Nevertheless, we model the two components separately for completeness. The shear-corrected effective radius for the NW source is $ 227_{-53}^{+55}$ pc, and an upper limit for the SE source is $138_{-62}^{+95}$ pc. 

\begin{figure} 
 \gridline{
 \fig{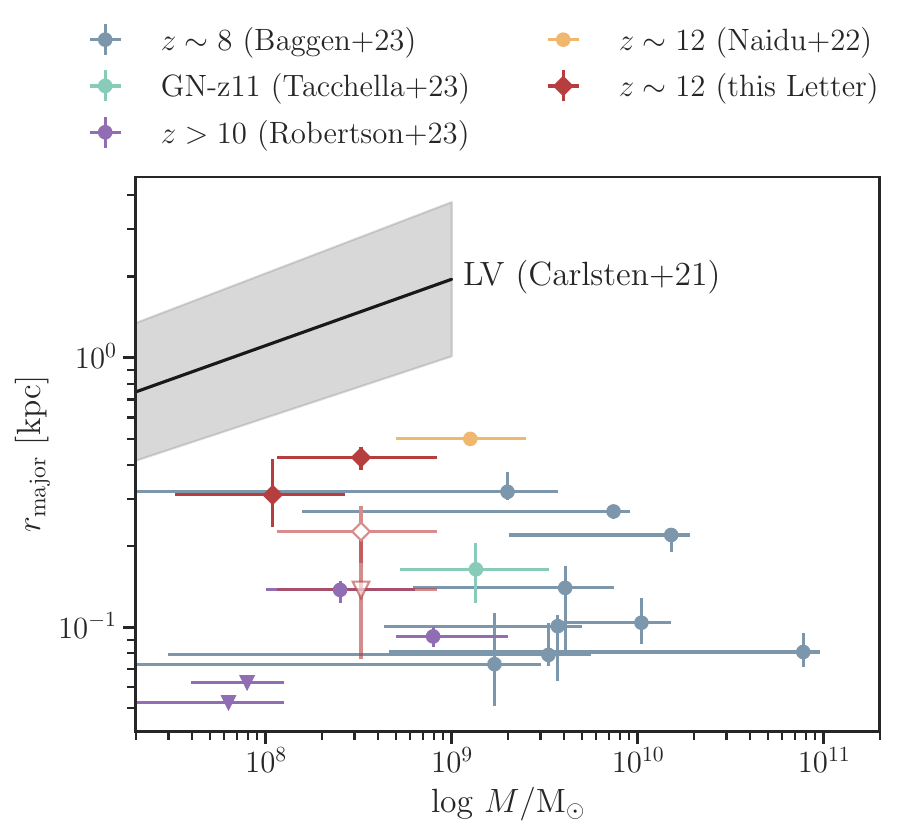}{0.46\textwidth}{}
 }
 \caption{Mass--size relationship. \twelve\ and \thirteen\ are shown as red diamonds. We take the average masses and uncertainties from the three SED fitting results. Also shown in unfilled red symbols are the \twelve\ sizes assuming two components. The triangle indicates an upper limit. We include five data sets from the literature as well: massive galaxies at $z\sim8$ \citep{Baggen2023,Labbe2023}, GN-z11 at $z \sim 11$ \citep{Tacchella2023:gnz11}, a plausible luminous galaxy at $z \sim 12$ \citep{Naidu2022:z12,Bakx2023}, and $z>10$ galaxies confirmed by JADES \citep{Curtis-lake2023,Robertson2023}. The gray line is a fit of the mass--size relationship for satellites of Milky Way-like hosts \citep{Carlsten2021}; note that the local volume (LV) relation is measured at different rest-frame wavelengths than the high-redshift samples, and is included here only for reference. The galaxies of this Letter are among the largest known at $z \gtrsim 10$.}
\label{fig:mass_size}
\end{figure}

We place the morphological measurements in the context of other high-redshift samples in Figure~\ref{fig:mass_size}. Interestingly, the galaxies of this Letter are notably larger than the $z>10$ galaxies reported in \citet{Robertson2023}, and are comparable to the massive galaxies at $z\sim8$ \citep{Baggen2023,Labbe2023}. These measurements may hint at a scatter in the mass--size relationship at early epochs, and that selection effects could bias samples toward the most compact objects in shallower surveys.

To conclude, the earliest galaxies are crucial in understanding the theories of galaxy and structure formation, yet only three $z > 12$ galaxies have been confirmed in the literature thus far \citep{Bakx2023,Curtis-lake2023,Robertson2023}. 
The discoveries presented here therefore provide valuable insights into a redshift space now accessible by JWST.
It is hopeful that future systematic searches and investigations in this observational frontier will fill in one of the last missing pieces in modern astrophysics: the nature of the first luminous sources and their large-scale impact.

\section*{Acknowledgments}

This work is based in part on observations made with the NASA/ESA/CSA James Webb Space Telescope. The data were obtained from the Mikulski Archive for Space Telescopes at the Space Telescope Science Institute, which is operated by the Association of Universities for Research in Astronomy, Inc. (AURA), under NASA contract NAS 5-03127 for JWST. These observations are associated with JWST-GO-2561, JWST-ERS-1324, and JWST-DD-2756. This research is also based on observations made with the NASA/ESA Hubble Space Telescope obtained from the Space Telescope Science Institute under NASA contract NAS 5–26555. These observations are associated with programs HST-GO-11689, HST-GO-13386, HST-GO/DD-13495, HST-GO-13389, HST-GO-15117, and HST-GO/DD-17231. The specific observations analyzed can be accessed via \dataset[10.17909/8k5c-xr27]{http://dx.doi.org/10.17909/8k5c-xr27}.

BW thanks Ben Johnson for discussions on the size measurements of the JADES galaxies. BW and JL acknowledge support from JWST-GO-02561.022-A. 
AZ and LJF acknowledge support by grant 2020750 from the United States-Israel Binational Science Foundation (BSF) and grant 2109066 from the United States National Science Foundation (NSF), and by the Ministry of Science \& Technology, Israel.
HA and IC acknowledge support from CNES, focused on the JWST mission, and the Programme National Cosmology and Galaxies (PNCG) of CNRS/INSU with INP and IN2P3, co-funded by CEA and CNES.
RB acknowledges support from the Research Corporation for Scientific Advancement (RCSA) Cottrell Scholar Award ID No.: 27587.
PD acknowledges support from the NWO grant 016.VIDI.189.162 (``ODIN") and the European Commission's and University of Groningen's CO-FUND Rosalind Franklin program.
JEG acknowledges NSF/AAG grant \#1007094.
YF acknowledges support from NAOJ ALMA Scientific Research grant No. 2020-16B, and from JSPS KAKENHI grant No. JP23K13149. 
RP and DM acknowledge support from JWST-GO-02561.013-A.
The work of CCW is supported by NOIRLab, which is managed by AURA under a cooperative agreement with the NSF.
The Cosmic Dawn Center is funded by the Danish National Research Foundation (DNRF) under grant \#140.

Computations for this research were performed on the Pennsylvania State University's Institute for Computational and Data Sciences' Roar supercomputer. This publication made use of the NASA Astrophysical Data System for bibliographic information.

\facilities{HST(ACS,WFC3), JWST(NIRCam, NIRSpec)}
\software{
Astropy \citep{2013A&A...558A..33A, 2018AJ....156..123A,2022ApJ...935..167A}, 
Bagpipes \citep{Carnall2018,Carnall2019:VANDELS}, 
BEAGLE \citep{Chevallard2016}, 
Cloudy \citep{Ferland:2017}, 
Corner \citep{2016JOSS....1...24F}, 
EAzY \citep{Brammer2008}, 
FSPS \citep{Conroy2010}, 
dynesty \citep{Speagle2020}, 
Matplotlib \citep{2007CSE.....9...90H}, 
msaexp \citep[][v0.6.10]{Brammer2022}, 
NumPy \citep{2020Natur.585..357H}, 
NUTS \citep{2011arXiv1111.4246H,2019arXiv191211554P}, 
Prospector \citep{Johnson2021}, 
PyMultinest \citep{Buchner:2014, Feroz:2019}, 
pysersic \citep{Pasha2023}, 
SciPy \citep{2020NatMe..17..261V}
}

\appendix
\counterwithin{figure}{section}
\counterwithin{table}{section}

\section{Parent sample\label{app:sample}}

We supplement further details on the parent sample. Table~\ref{tab:failed} lists all the candidates that show photometric redshifts $>12$, but were not confirmed to be high-redshift after the spectroscopic observations due to various reasons also noted in the table. Figure~\ref{fig:app:balmer} includes all the spectra, accompanied by $p(z)$.

Out of a total of eight candidates, three turn out to be too faint to infer the nature of the sources. One candidate may be affected by the modeling of the brightest cluster galaxies in the vicinity and has been deemed as an artifact in later iterations of the photometric reduction pipeline. For the remaining two failed candidates, one exhibits a plausible emission line which leads us to favor a low-redshift solution, whereas the other candidate is a textbook example of confusion between Lyman and Balmer breaks. The S/N of the mean flux red- and blue-ward of the break are 4.77 and 2.65, respectively, corroborating the Balmer break diagnostic.

It is also interesting to revisit the pre-spectra photometric redshifts. \eazy\ identifies all 8 candidates at $z>11$. Medians of the redshift posteriors from \prospector\ places one of them at low redshift, whereas the maximum-likelihood redshifts indicate 4 of them are better fitted with a low-redshift model. The latter suggests that it is possible to improve the photometric inference via more informative priors. The effect of including a number density prior, as proposed in \citealt{Wang2023:pbeta}, will be explored in the next generation of stellar population catalogs as part of the UNCOVER survey.

Finally, a word of caution: it is too early to draw conclusions about selection effects or success rates; here we merely aim to paint a more complete picture of the sample.

\section{Forced fits at low redshift\label{app:balmer}}

As an additional check, we fit the observed spectra adapting the same procedure outlined in Section~\ref{subsec:data:zspec}, but restricting the redshift range to be $0 < z < 10$. The best-fit model spectra and $p(z)$ are shown in Figure~\ref{fig:app:data}.

We also compare the $\chi^2$ values between the best-fit models from \bagpipes\ at high and low redshifts. Similar to the findings above, the $z > 12$ solutions are preferred for both sources.

\renewcommand{\thetable}{A.1}
\begin{deluxetable*}{lcccccl}
\tablecaption{Low-redshift Interlopers or Unconfirmed Sources\label{tab:failed}}
\tablehead{
\colhead{ID} & \colhead{RA [deg]} & \colhead{Dec [deg]} & \colhead{$z_{\rm eazy}$\tablenotemark{\scriptsize{a}}} & \colhead{$z_{\rm med, sps}$\tablenotemark{\scriptsize{b}}} & \colhead{$z_{\rm ml, sps}$\tablenotemark{\scriptsize{c}}} & \colhead{Notes}}
\startdata
13805 & 3.56699 & -30.39996 & $11.38^{+0.28}_{-8.45}$ & $11.58^{+0.54}_{-9.78}$ & 2.21 & low-$z$ emission line \\
22002 & 3.63043 & -30.38362 & $11.42^{+1.67}_{-7.85}$ & $13.55^{+2.59}_{-9.42}$ & 3.38 & low S/N \\
40411 & 3.57991 & -30.35699 & $15.20^{+0.29}_{-0.75}$ & $15.81^{+0.44}_{-0.50}$ & 16.28 & data quality issue \\
44644 & 3.58291 & -30.34601 & $14.18^{+0.28}_{-2.14}$ & $14.24^{+0.91}_{-1.74}$ & 15.01 & Balmer break \\
45390 & 3.55640 & -30.34433 & $15.68^{+0.56}_{-11.01}$ & $15.36^{+1.17}_{-11.22}$ & 4.88 & low S/N \\
51704 & 3.55716 & -30.32870 & $11.63^{+0.21}_{-8.88}$ & $1.86^{+0.54}_{-9.78}$ & 3.02 & low S/N \\
\enddata
\tablenotetext{a}{Photometric redshift from \eazy{}.}
\tablenotetext{b}{Posterior median of photometric redshift from \prospector{}.}
\tablenotetext{c}{Maximum-likelihood photometric redshift from \prospector{}.}
\end{deluxetable*}

\renewcommand{\thefigure}{A.1}
\begin{figure*} 
\gridline{
  \fig{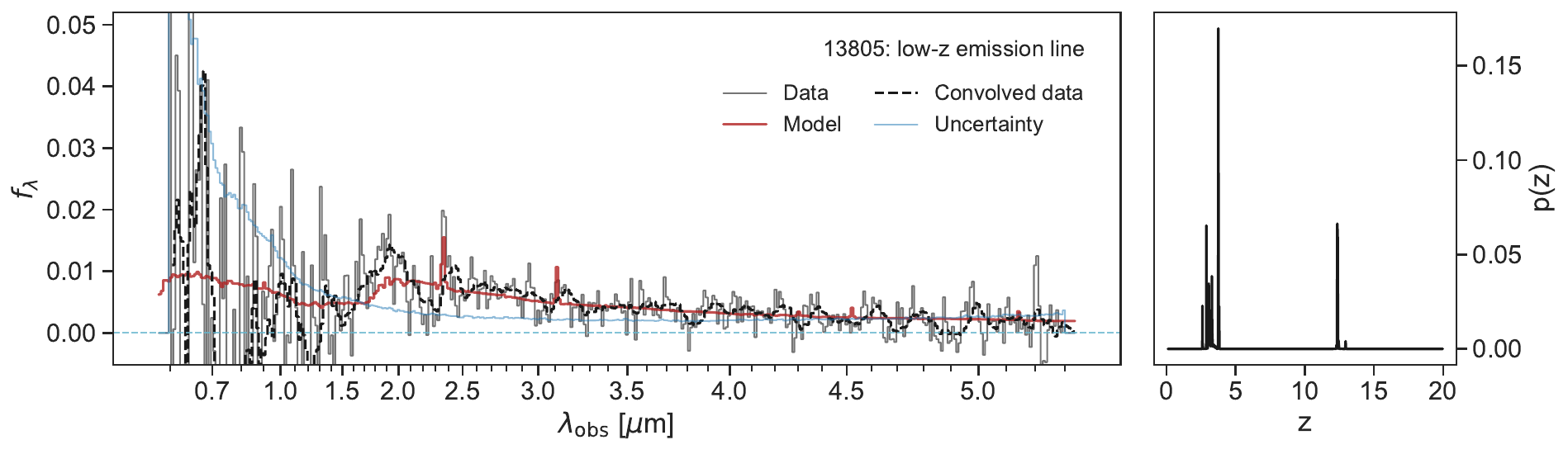}{0.95\textwidth}{}
}
\gridline{
  \fig{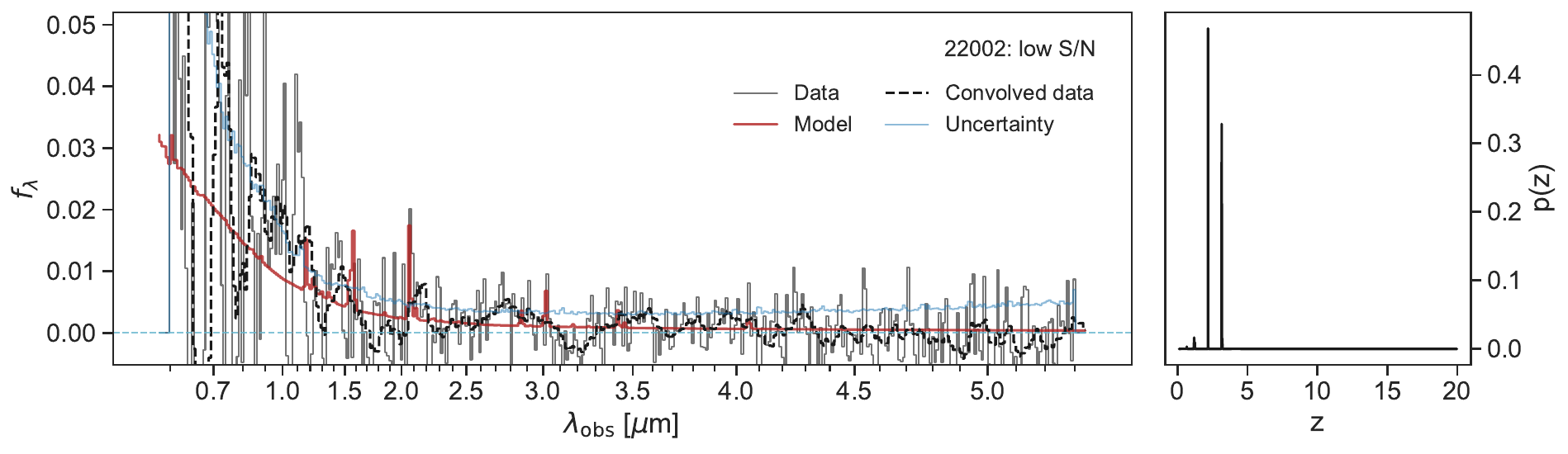}{0.95\textwidth}{}
}
\gridline{
  \fig{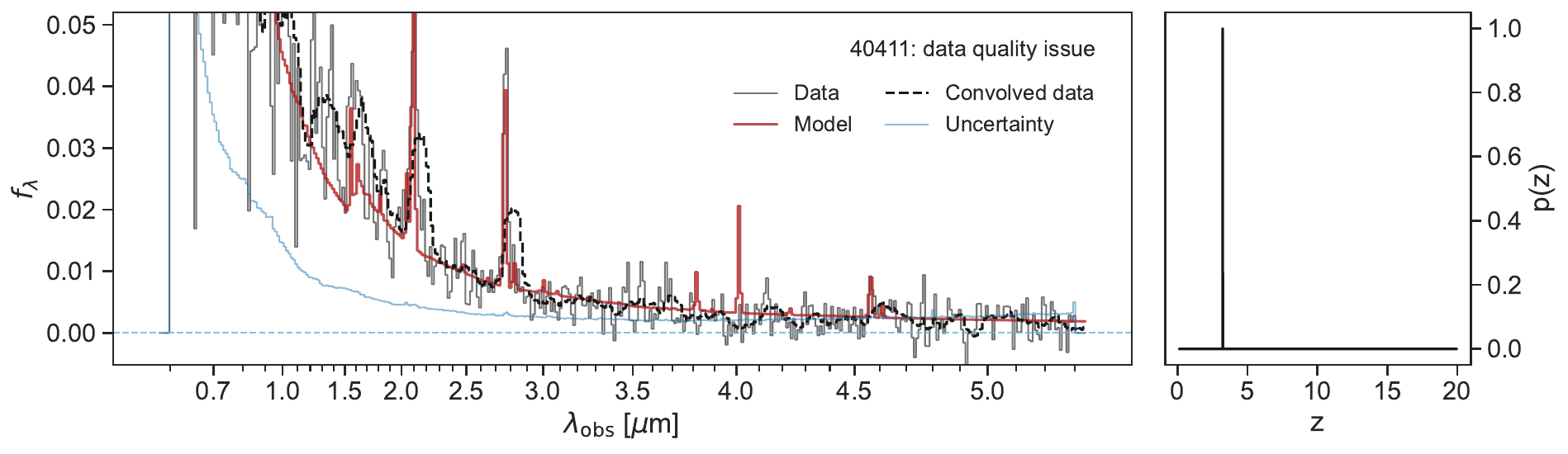}{0.95\textwidth}{}
}
\caption{Low-redshift interlopers or unconfirmed sources. For each figure, the left panel displays the 1D spectrum in $f_\lambda$ as a function of the observed wavelength. Data, the best-fit \eazy\ model, and uncertainties are plotted in gray, red, and blue, respectively. A convolved observed spectrum is additionally shown as a dashed black line for visualization purposes. The cyan horizontal line is at $y=0$ to guide the eye. The right panel shows the probability density over the redshift range of the search for a minimum $\chi^2$.
}
\label{fig:app:balmer}
\end{figure*}

\renewcommand{\thefigure}{A.1}
\begin{figure*} 
\gridline{
  \fig{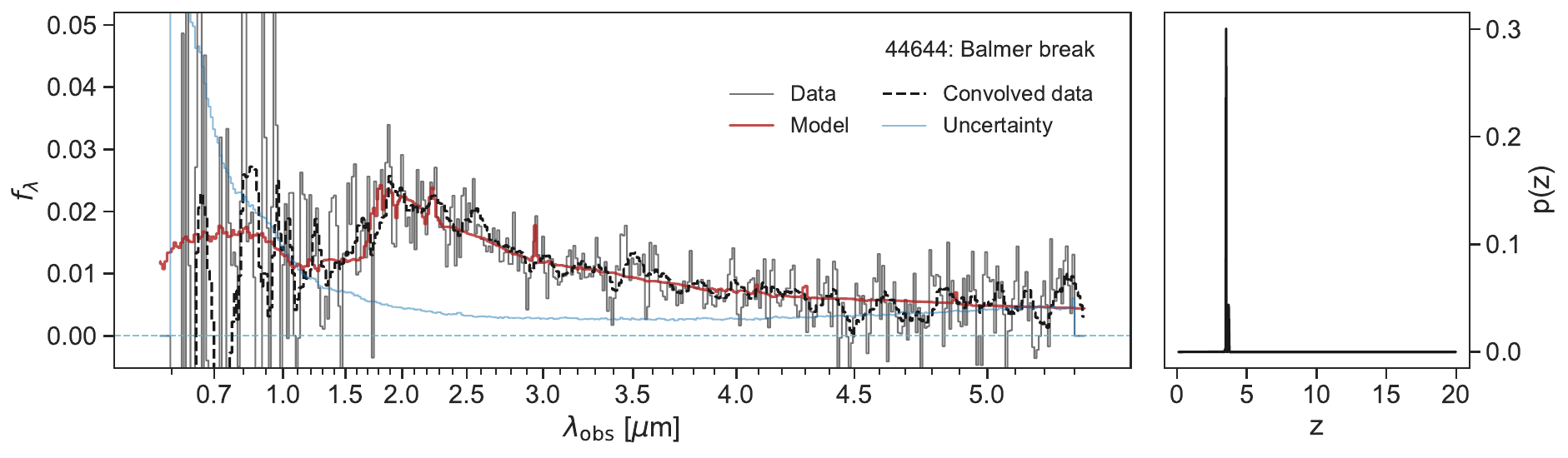}{0.95\textwidth}{}
}
\gridline{
  \fig{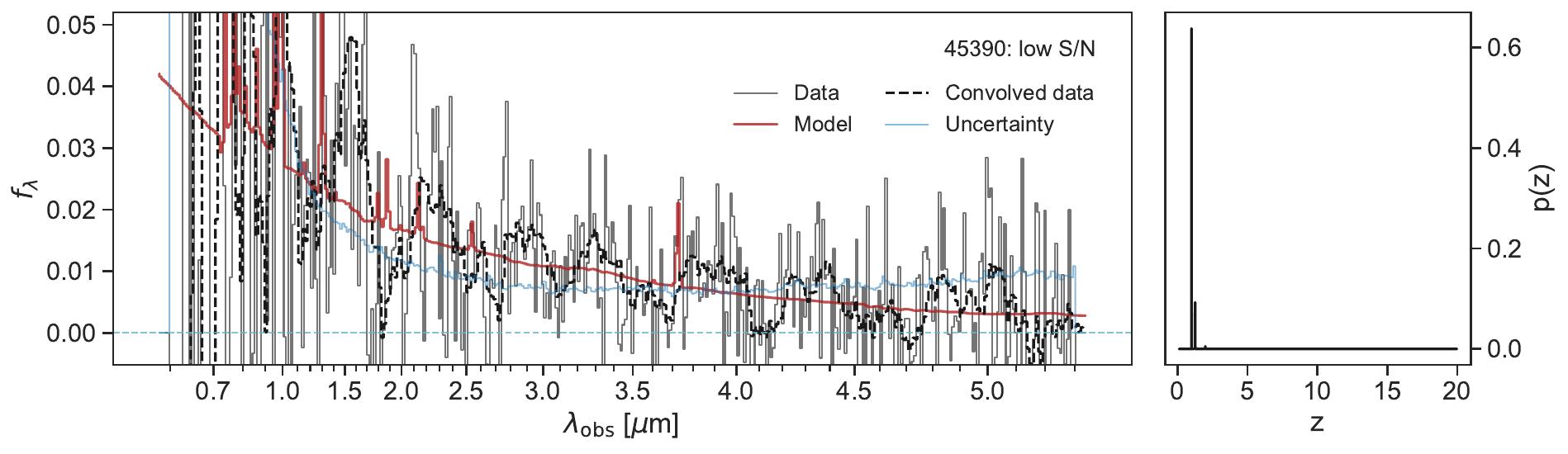}{0.95\textwidth}{}
}
\gridline{
  \fig{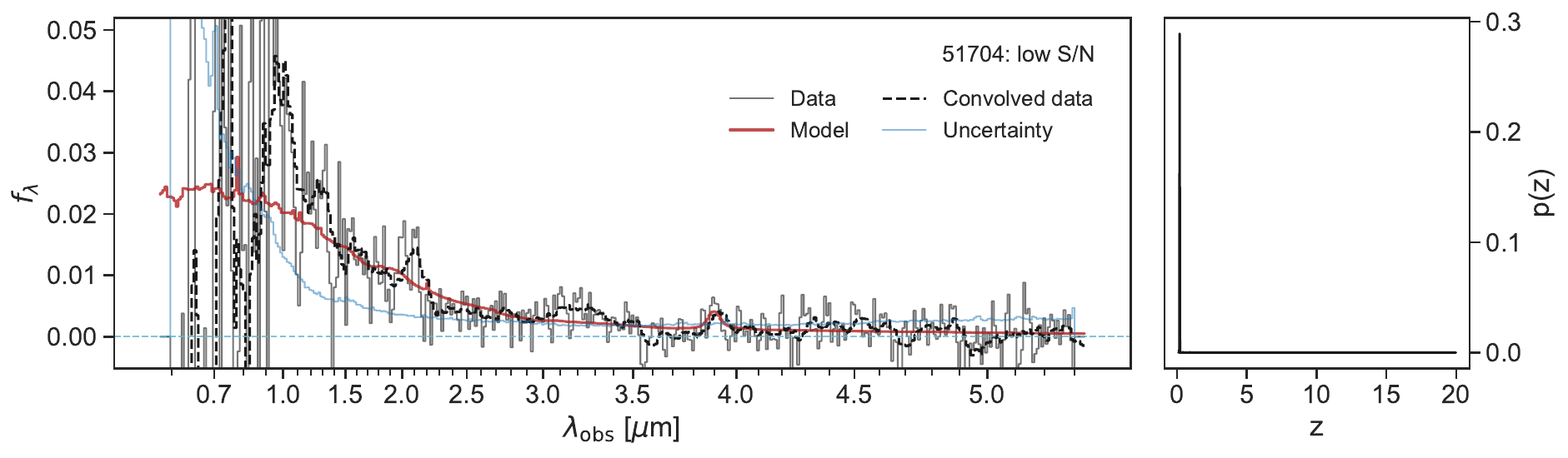}{0.95\textwidth}{}
}
\caption{(continued.)}
\end{figure*}

\renewcommand{\thefigure}{B.1}
\begin{figure*} 
 \gridline{
 \fig{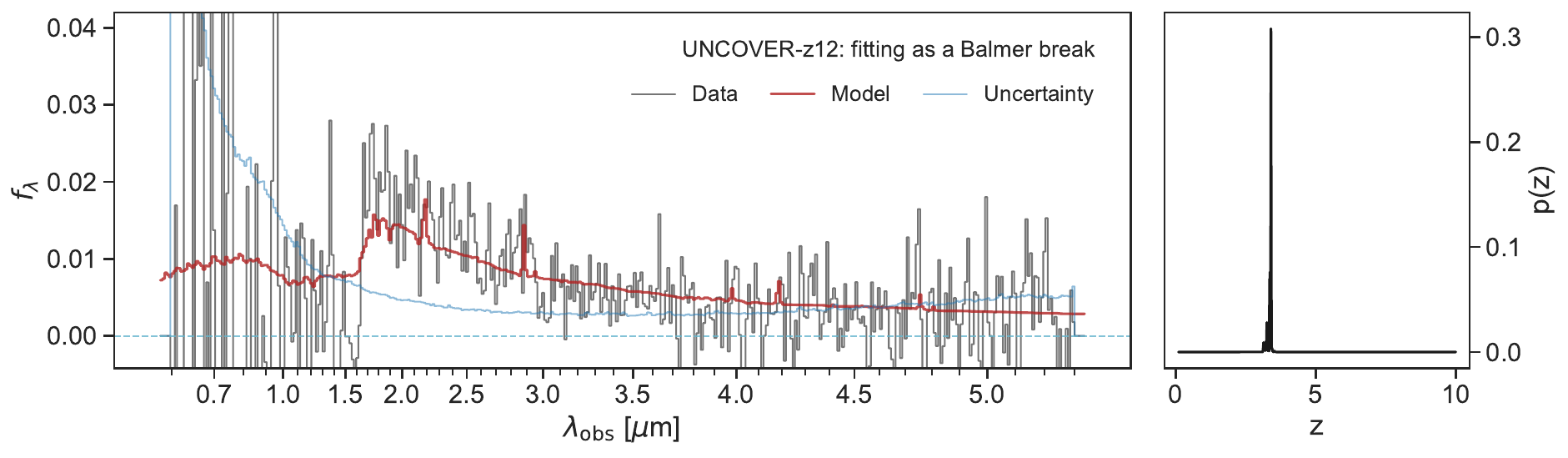}{0.95\textwidth}{(a)}
 }
 \gridline{
 \fig{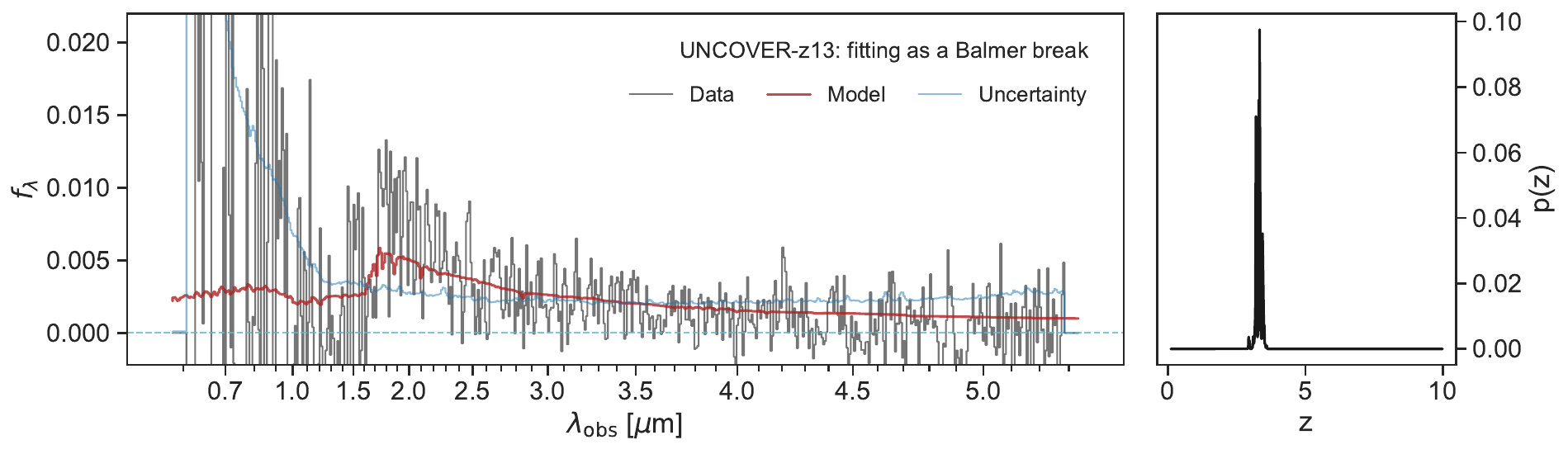}{0.95\textwidth}{(b)}
 } 
\caption{Forced fits at low redshift. 
(a) The left panel displays the 1D spectrum for \twelve\ in $f_\lambda$ as a function of the observed wavelength. Data are plotted in gray, whereas the best-fit \eazy\ model at $z \sim 3$ is plotted in red. Also plotted are the uncertainties in blue. The cyan horizontal line is at $y=0$ to guide the eye. The right panel shows the probability density over the redshift range of the search for a minimum $\chi^2$. (b) Same as the above figure, but for \thirteen. The high-redshift solutions are favored for both sources.}
\label{fig:app:data}
\end{figure*}

\bibliography{uncover_zspec_wang.bib}

\end{document}